\documentclass[twocolumn,
aps,prl,twoside,superscriptaddress,floatfix
,10pt,tightenlines
,footinbib
,notitlepage
]{revtex4-1}

\usepackage{mathtools}
\usepackage{dcolumn}
\usepackage{float}
\usepackage{bm}
\usepackage[T1]{fontenc}
\usepackage{amsmath}
\usepackage{graphicx} 
\usepackage{tabularx}
\usepackage{multirow}
\usepackage{rotating} 
\usepackage{makecell}
\usepackage{hyperref}
\hypersetup{colorlinks=true, linkcolor=blue, citecolor=blue, urlcolor=blue}
\usepackage{tikz}
\usepackage[normalem]{ulem}
\usepackage{pgfplots}

\begin{document}

\title{
Fractional-flux oscillations of Josephson critical currents\\ in multi-gap superconductors: a test for unconventional superconductivity.
}
\author{M. Donaire}
\affiliation{Departamento de Física Teórica, Atómica y Óptica and IMUVA, Universidad de Valladolid,
Paseo Belén 7, 47011 Valladolid, Spain}
\author{A. Cano}
\affiliation{Univ. Grenoble Alpes, CNRS, Grenoble INP, Institut Néel, 25 Rue des Martyrs, 38042, Grenoble, France
}
\date{\today}

\begin{abstract}
Josephson-junction interferometry has played a pivotal role in uncovering unconventional superconductivity in the cuprates. Using a Ginzburg-Landau-like approach, 
we generalize previous results to the genuine multi-gap case. Thus, we show that fractional flux oscillations of the Josephson critical current can arise as a direct consequence of multi-gap superconductivity. 
These oscillations reveal key information about the underlying superconducting states, including the unconventional $s_\pm$-wave state. Thus, our findings suggest new phase-sensitive experiments to characterize the Cooper pairing of new emerging superconductors such as the nickelates.
\end{abstract}

\maketitle

Superconducting nickelates stimulate new research perspectives in condensed-matter physics. The discovery of superconductivity in these systems \cite{hwang19a,mundy21,sun23,li24} is difficult to reconcile with the conventional electron-phonon coupling mechanism \cite{arita19,held23,meier24,Ouyang2024,You2025}. 
Thus, nickelates are widely believed to host unconventional superconductivity, and this is believed to happen in two distinct forms. 
In the infinite-layer case, the superconducting state is generally considered as essentially analogous to the single-gap $d$-wave state of the cuprates (see e.g. \cite{arita19,thomale20,held20,held24}). In the Ruddlesden-Popper case, in contrast, a multi-gap $s_\pm$-wave state has been argued to be a very likely state, thus putting these systems in another category where the iron-based superconductors are the reference \cite{kuroki17,kuroki24}. 
Experimentally, however, the unambiguous determination of these states is challenging and the question remains open. In this respect, phase-sensitive tests based in the Josephson effect arguably provide the most elegant and direct methods for resolving this question. 

In this paper, we analyze the behavior of simple Josephson-junction (JJ) schemes that can be used for attempting such tests. Specifically, we consider the JJ coupling between two different superconductors and examine the response as a function of the magnetic flux. 
We consider, in particular, genuine multi-gap cases in which the amplitude of the corresponding gaps is different. Thus, find distinct responses in which the Josephson critical current exhibit fractional-flux oscillations. These oscillations encode information about the underlying superconducting state, and hence can reveal its precise nature.   
To emphasize the universality of our findings as well as their novelty compared with previous ones, we choose the Ginzburg-Landau-like approach to conduct our analysis.

\begin{figure}[b!]
    \centering
    \includegraphics[width=0.485\textwidth]{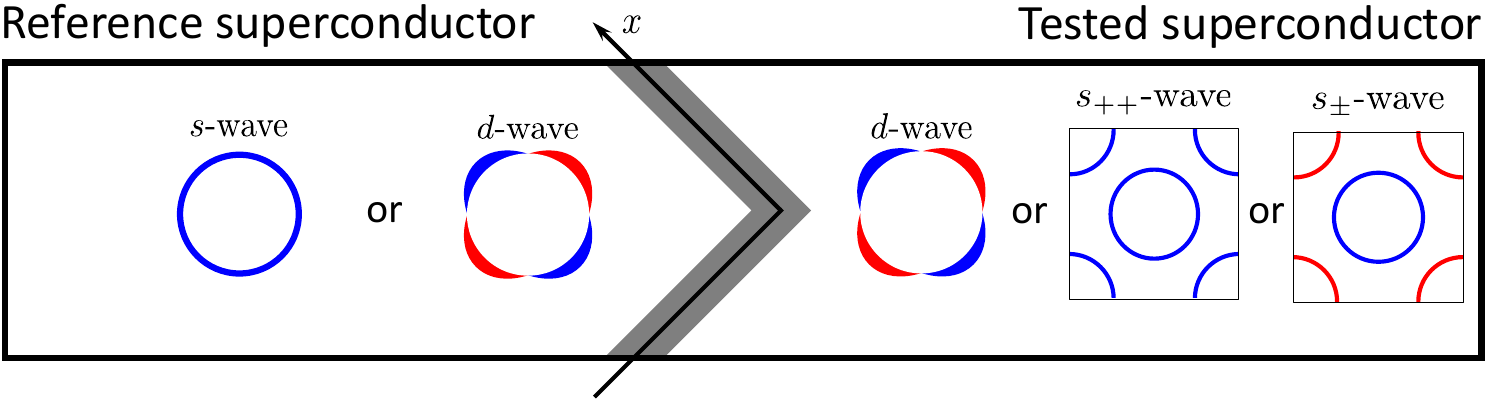}
    \caption{Sketch of a extended Josephson junction in a corner geometry. We consider single-gap reference superconductors that can be either conventional ($s$-wave) or unconventional ($d$-wave) to test the state of another superconductor.  
    }
    \label{fig:JJ}
\end{figure}

We first recall well-established results for single-gap superconductors. Consider the extended JJ sketched in Fig. \ref{fig:JJ}. The Josephson current density is $j(x) = j_{c} \sin \varphi(x)$, where $j_c$ is the critical current and  $\varphi(x)$ denotes the gauge-invariant phase difference between the superconductors at the point $x$ along the JJ (see e.g. \cite{harlingen95-rmp,buckel2008}). 
In the presence of a magnetic flux $\Phi$, the variation of $\varphi$ along the JJ is such that
\begin{align}
    {d\varphi \over dx} = 2 \pi \frac{\Phi}{\Phi_0}\frac{1}{l}, 
\end{align}
where $\Phi_0$ is the flux quantum and $l$ the total length of the JJ. 
In general, $\varphi$ can be taken as
\begin{align}
    \varphi (x) = \varphi_0 + 2 \pi \frac{\Phi}{\Phi_0}\frac{x}{l} + \varphi_*(x),
\end{align}
where $\varphi_0$ is a reference phase and $\varphi_*(x)$ describes the ``geometrical'' phase shift engineered in the setup. 
Without loss of generality, we assume that the JJ is made of two parts of the same length. Further, it is convenient to take $\varphi_*$ as $-{\delta/2}$ in one part ($-{l/2}<x<0$) and ${\delta/2}$ in the other ($0<x<{l/2}$). Thus, integrating the current along the JJ one obtains 
\begin{align}
I =  
I_c \frac{\sin (\pi \tfrac{ \Phi}{2\Phi_0})\cos (\pi \tfrac{ \Phi}{2\Phi_0} + \frac{\delta}{2})}{ \pi \frac{ \Phi}{2 \Phi_0}} \sin \varphi_0.
\label{I-single-gap}\end{align}

\begin{figure}[t!]
    \centering
    \includegraphics[width=0.425\textwidth]{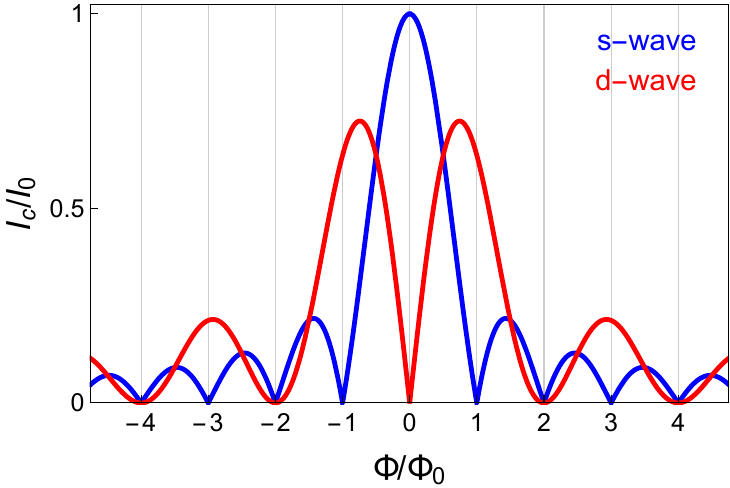}
    \caption{
    Fraunhofer diffraction patterns for the modulation of the critical current with the applied magnetic field in corner JJs. 
    }
    \label{fig:fraunhofer-same}
\end{figure}

In the conventional $s$-wave case the geometrical phase shift is always $\delta = 0$. However, this shift can be nonzero for unconventional superconductors depending on the specific configuration of the setup. In particular, $\delta$ can be engineered to be $\pi$ for a corner JJ between a $s$-wave superconductor and a $d$-wave one if the crystal axes of the latter are properly aligned. The difference between the corresponding Fraunhofer diffraction patterns is illustrated in Fig. \ref{fig:fraunhofer-same}.

From such distinct Fraunhofer patterns, it was possible to conclude the $d$-wave nature of the superconducting cuprates (see e.g. \cite{harlingen95-rmp,buckel2008}). Thus, it would be natural to attempt similar experiments to probe the nature of the superconducting state of the nickelates. These experiments could thus provide direct evidence of the $d$-wave superconducting state hypothesized for some of these materials. 

To obtain direct evidence of unconventional $s_\pm$-wave superconductivity, however, is more challenging. Similarly to the $d$-wave, the $s_\pm$ state is defined by a superconducting gap function displaying a $\pi$-phase shift between its different parts, say $\Delta_{1} = |\Delta_{1}|e^{i\phi_{1}}$ and $\Delta_{2} = |\Delta_{2}|e^{i\phi_{2}}$ where $\phi_2 - \phi_1 = \pi$. However, contrary to the $d$-wave case, these parts are related to different electronic bands---rather than to the same band as in the cuprates. Consequently, there is no straightforward relation with the crystal axes. Nevertheless, several proposals have been discussed to probe $s_\pm$-wave state in the case of the iron-based superconductors (see e.g. \cite{seidel11}). In particular, the JJ tunneling could be engineered to (locally) favor the current from one band or the other and thus to obtain a $\pi$-phase shift. This could be achieved by means of the variation of the JJ thickness and/or the use of doping to control the relative alignment of the corresponding bands \cite{mazin09-prl,phillips09-prb}, which could be easily borrowed for the nickelates. 

To gain further insight in this direction, we revisit the theory for the multi-gap case. 
For the sake of simplicity, we consider a two-gap superconductor in which the current density is
$\mathbf{j}=
i K_1
(\Delta^*_1\nabla \Delta_1 - \Delta_1 \nabla \Delta^*_1)
+
i K_2 
(\Delta^*_2\nabla \Delta_2 - \Delta_2 \nabla \Delta^*_2)
 - 2\big(K_1|\Delta_1|^2  + K_2 |\Delta_2|^2 \big)\big(\frac{2\pi}{\Phi_0}\big) \mathbf{A}, 
$
where $K_{1,2}$ are the gradient coefficients of the corresponding Ginzburg-Landau functional. Further, we consider the same JJ setup as before to test the state of this superconductor (see Fig. \ref{fig:JJ}). 
The Josephson current density then becomes $j(x) = j_{c1} \sin \varphi_1(x) + j_{c2} \sin \varphi_2(x)$, where $\varphi_i = \phi_i - \phi_0$ denotes the phase difference of each gap with respect to 
the one ($\phi_0$) of the single-gap reference superconductor. 
In this case, the variation of these phases along the JJ is such that
\begin{align}
\nu_1 {d \varphi_1 \over dx} 
+
\nu_2 {d \varphi_2 \over dx}
&= 
2  
\pi {\Phi\over \Phi_0}{1\over l},
\label{relation-phases} \end{align}
where 
$ \nu_{i} = 
K_i|\Delta_i|^2 /(
{ K_1|\Delta_1|^2 + K_2|\Delta_2|^2 })$ (see e.g. \cite{ota09-prl}).  

Eq. \eqref{relation-phases} admits multiple solutions. A simple one is 
\begin{align}
\varphi_{i} 
&=
\varphi_{0i}
+ 
2 \pi {\Phi\over \Phi_0}{x \over l}, 
\label{same-gap} 
\end{align}
where $\varphi_{0i}$ are reference phases whose difference is either $0$ ($s_{++}$-wave) or $\pi$ ($s_\pm$-wave). Thus, the integral of the Josephson current density yields
\begin{align}
I = 2 (I_{c1} \pm I_{c2} )
\frac{\sin^2 (\pi \tfrac{ \Phi}{\Phi_0})}
{ \pi \frac{ \Phi}{\Phi_0}}
\sin\varphi_{0}.
\label{fraunhofer-same}
\end{align}
This result coincides with the one reported in \cite{ota09-prl} (see also \cite{sperstad09-prb}), where the reduction and eventually vanishing of the amplitude of Fraunhofer diffraction pattern has been proposed as a test for $s_\pm$-wave superconductivity. 

At this point, we notice the following. The solution obtained according to Eq. \eqref{same-gap} captures the 0 or $\pi$ difference between the phases of the two gaps by construction. However, these gaps are tacitly assumed to be equivalent otherwise. Indeed, the parametrization \eqref{same-gap} fails to reproduce the single-gap limit in which the amplitude of one of the gaps tends to zero, so that the magnetic field is solely screened by the supercurrents associated with the other one. This means that the result obtained from \eqref{fraunhofer-same} in limit $I_{c2} \to 0$ (or $I_{c1} \to 0$) needs to be understood as resulting from a ``one-channel'' tunneling across the JJ rather than due to the effective $\Delta_2 \to 0$ (or $\Delta_{1} \to 0$) single-gap behavior of the superconductor itself. 

\begin{figure*}[t!]
    \centering
    \includegraphics[width=0.425\textwidth]{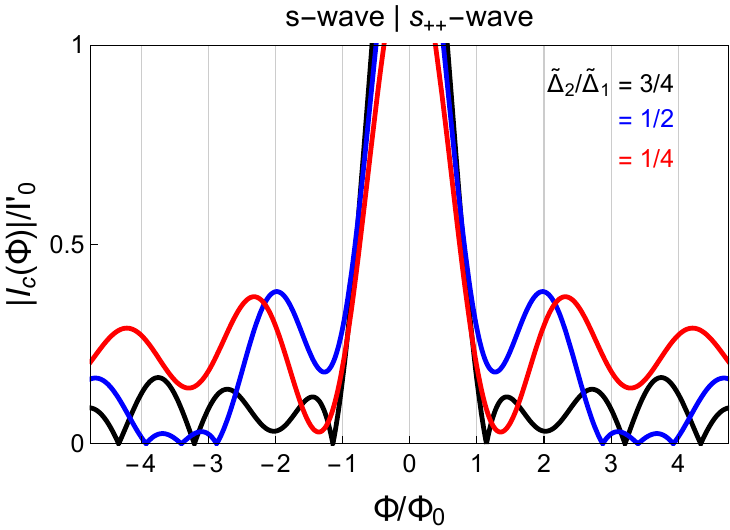}
    \hspace{5em}
    \includegraphics[width=0.425\textwidth]{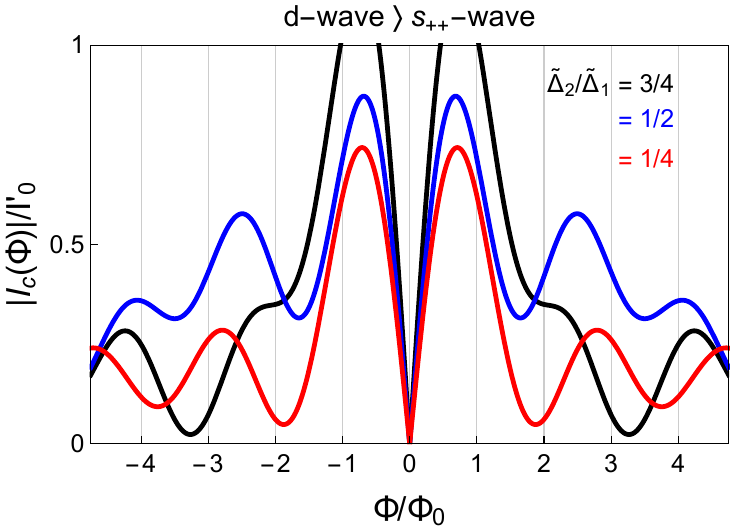}
    \\[2em]
    \includegraphics[width=0.425\textwidth]{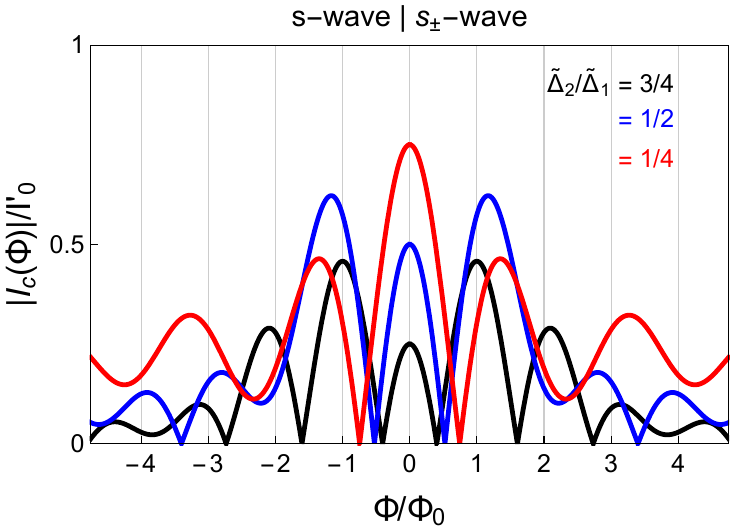}
    \hspace{5em}
    \includegraphics[width=0.425\textwidth]{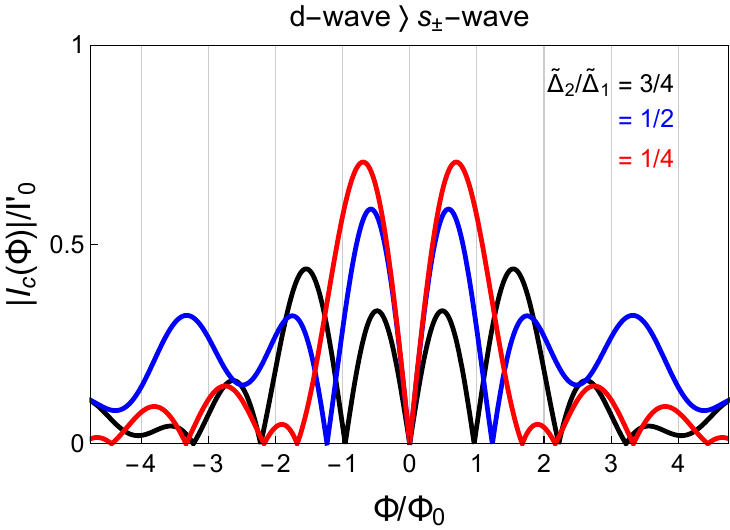}
    \caption{
    Fraunhofer diffraction patterns for the modulation of the critical current with the applied magnetic field for JJs between single-gap and two-gap superconductors (to be compared with Fig. \ref{fig:fraunhofer-same}). The right panels correspond to a 45$^{\circ}$-corner JJs where the crystal axes of the reference $d$-wave superconductor are properly aligned. 
    The individual critical-current parameters $I_{ci}$ are assumed to be $I_0' |\Delta_i|$. Due to the different superconducting-gap amplitudes $|\Delta_i|$, the modulation of the individual Josephson currents display different non-universal temperature-dependent periods. Thus, overall modulation then becomes aperiodic and hence incommensurate with the flux quantum in both $s_{++}$- and $s_\pm$-wave cases.   
    }
    \label{fig:fraunhofer-different}
\end{figure*}

We find that a more general solution of Eq. \eqref{relation-phases} is   
\begin{align}
\varphi_{i} 
&=
\varphi_{0i}
+ 
2 \pi r_i{\Phi\over \Phi_0}{x \over l}
+ \varphi_{*i}(x),
\label{different-gap} 
\end{align}
where the parameters $r_i$ satisfy the condition $\nu_1 r_1 + \nu_2 r_2 =1$ 
and  
$\varphi_{*i}(x)$ describes the geometrically-engineered phase shift for each gap. 
Further, it is convenient to take this latter as $\varphi_{*i} =-\delta/2$ if $-l/2<x<0$ and $\delta/2$ if $0<x<l/2$ for the $s_{++}$-wave case, and $\varphi_{*1,2} =\mp \delta/2$ if $-l/2<x<0$ and $\pm \delta/2$ if $0<x<l/2$ for the $s_\pm$-wave one.  
Thus, we obtain the total current
\begin{align}
I = 
\Bigg(
& I_{c1}
\frac{
 \sin \big(
\pi r_1 
\frac{\Phi}{ 2\Phi_0} \big) \cos \big(
\pi r_1 
\frac{\Phi}{ 2\Phi_0} + \frac{\delta}{2}\big)
}
{\pi r_1 \frac{\Phi}{ 2\Phi_0} }
\nonumber \\
& 
\pm 
I_{c2} \frac{
\sin \big(\pi r_2 
\frac{\Phi}{ 2\Phi_0}\big)
\cos \big(\pi r_2 
\frac{\Phi}{ 2\Phi_0} -\frac{\delta}{2}\big)
}
{\pi r_2 \frac{\Phi}{ 2\Phi_0} }
\Bigg)\sin\varphi_{0},
\label{fraunhofer-different}
\end{align}
where $+$ and $-$ refers to the conventional $s_{++}$-wave and unconventional $s_\pm$-wave states of the two-gap superconductor respectively. 

From Eq. \eqref{fraunhofer-different}, different JJ geometries of interest can be described via the phase shift $\delta$. As in the entirely single-gap case, the geometry of a straight JJ between the reference single-gap superconductor, either $s$-wave or $d$-wave, and the two-gap superconductor corresponds to $\delta = 0$. 
In that case, Eq. \eqref{fraunhofer-different} reduces to
\begin{align}
I = 
\Bigg(
& I_{c1}
\frac{
 \sin \big(
\pi r_1 
\frac{\Phi}{\Phi_0}\big)
}
{
\pi r_1 
\frac{\Phi}{\Phi_0} }
\pm
I_{c2} \frac{
\sin \big(
\pi r_2 
\frac{\Phi}{ \Phi_0}\big)
}
{\pi r_2 
\frac{\Phi}{\Phi_0} }
\Bigg)\sin\varphi_{0}.
\label{fraunhofer-different-s}
\end{align}
In contrast, when the reference superconductor is $d$-wave, the case of a 45$^\circ$-corner JJ corresponds to $\delta = \pi$. In this configuration, the Josephson current becomes
\begin{align}
I = 
\Bigg(
& I_{c1}
\frac{
 \sin^2 \big(
\pi r_1 
\frac{\Phi}{2\Phi_0}\big)
}
{
\pi r_1 
\frac{\Phi}{2\Phi_0} }
\pm
I_{c2} \frac{
\sin^2 \big(
\pi r_2 
\frac{\Phi}{ 2\Phi_0}\big)
}
{\pi r_2 
\frac{\Phi}{2\Phi_0} }
\Bigg)\sin\varphi_{0}.
\label{fraunhofer-different-d}
\end{align}

To illustrate these results, we chose the particular solution $r_i = \nu_i /(\nu^2_1 + \nu^2_2)$ since the physics is quite transparent in this case \footnote{
More general solutions can be parametrized as 
$r_1 = 1 + \sqrt{\frac{K_2 |\Delta_2|^2}{K_1 |\Delta_1|^2}} r$ 
and 
$r_2 = 1 - \sqrt{\frac{K_1 |\Delta_1|^2}{K_2 |\Delta_2|^2}} r$, 
where $r$ may depend on the flux. 
The particular solution $r_i = \nu_i /(\nu^2_1 + \nu^2_2)$ corresponds to $r = \frac{(K_1 |\Delta_1|^2-K_2 |\Delta_2|^2)\sqrt{K_1 K_2 |\Delta_1|^2|\Delta_2|^2}}{K_1^2 |\Delta_1|^4 + K_1^2 |\Delta_1|^4}$
}. 
With this choice, the expression \eqref{fraunhofer-different} reproduces the single-gap result \eqref{I-single-gap} in the limit $|\Delta_2|/|\Delta_1| \to 0$ where $r_1 \to 1$ and $r_2 \to 0$ (or vice versa). 
In addition, for a straight JJ with no corner ($\delta =0$), this expression also reproduces the previous result Eq. \eqref{fraunhofer-same} in the limit $K_2^{1/2}|\Delta_2|/(K_1^{1/2}|\Delta_1|) \to 1$ (i.e. $r_{1,2 } \to {1}$) where the two gaps are equivalent. 
Thus, with $r_i = \nu_i /(\nu^2_1 + \nu^2_2)$, Eq. \eqref{fraunhofer-different} interpolates these limiting situations describing a general case in which the amplitude of the two superconducting gaps is different ($|\Delta_1| \neq |\Delta_2|$).

The corresponding Fraunhofer patterns are illustrated in Fig. \ref{fig:fraunhofer-different} for different ratios of the gap amplitudes. To understand these patterns, notice that they originate from individual magnetic-field modulations with different periods $\Phi_0 / r_i $ that reflect the different amplitudes of the corresponding gaps ($r_i \propto |\Delta_i|^{2}$). These periods are therefore a non-universal fraction of the magnetic-flux quantum, which in addition will depend on temperature. This further leads to overall modulations of the critical current that become aperiodic and hence incommensurate with the flux quantum.

Interestingly, for specific values of the magnetic flux, $\Phi_{n,i} = n \Phi_0 / r_i$ ($n = \pm 1, \pm 2, ...$), the current is exclusively carried by one of the two gaps. This formally defines the multi-gap JJ as a quantum device with two operational states that can be switched via the magnetic flux. Further, in the $s_{\pm}$-wave case, switching between these two states could be used to change the phase difference across a subsequent JJ from 0 to $\pi$.          

The generalization to more than two gaps is straightforward and leads to essentially the same behavior.
Thus, we propose to perform JJ-interferometry experiments to test the superconducting state of materials such as the nickelates, where the observation of these distinct oscillations in the Josephson currents would probe the multi-gap nature of such a state. 
In fact, Fig. \ref{fig:fraunhofer-different} illustrates that the pattern of these oscillations is different in the conventional $s_{++}$ state compared to the unconventional $s_\pm$ one. Consequently, these experiments can in principle distinguish between these two fundamentally different cases.    

We notice that the above overall behavior can be viewed as the JJ version of the fractional-flux vortices recently reported for an iron-based superconductor \cite{babaev02-prl,moler23-science}. We also notice that, formally, Eq. \eqref{relation-phases} has other interesting solutions, among which we find solutions that are topologically different from \eqref{different-gap}. 
For example, $\varphi_{i} 
=
\varphi_{0i}
+ 
(-1)^{i}{2 \pi \over \nu_2 - \nu_1} {\Phi\over \Phi_0}{x \over l}$, which can be viewed as a neutral superflow state that gets unbalanced by the magnetic flux. This suggests the possibility of new types of states---including new Josephson vortices and textures---that can appear due to the multi-gap nature of the superconductor. 
Additional exotic solutions of this kind, however, generally become unstable in either the single-gap limit, the multi-gap but equal-amplitude limit, or both. 
To determine their relative stability in the general case is out of the scope of the present paper. In any case, we note that their emergence would also lead to fractional-flux oscillations of the Josephson critical currents, which is the main finding of our work. 

In summary, we have demonstrated that fractional-flux oscillations in Josephson critical currents can generally appear in multi-gap superconductors. 
These oscillations encode information about the superconducting gap amplitudes and symmetry, enabling the distinction between different types of superconducting states, including the unconventional $s_{\pm}$-wave one.  
Our findings thus provide a new basis for phase-sensitive experiments that is expected to motivate further work. In particular, the experimental realization of the proposed setups has the potential to provide critical insights into the pairing mechanisms of new superconducting materials such as the nickelates.

\bibliography{bib.bib}

\end{document}